\documentclass{emulateapj}
\usepackage{aas_macros}
\usepackage{color}
\begin{document}
\title{GLOBAL SPIRAL ARMS FORMATION BY NON-LINEAR INTERACTION OF WAKELETS}
\author{Jun Kumamoto and Masafumi Noguchi}
\affil{Astronomical Institute, Tohoku University, Aobaku, Sendai 980-8578, Japan}
\email{}
%

\begin{abstract}

The formation and evolution of galactic spiral arms is not yet clearly understood despite many
 analytic and numerical work.
Recently, a new idea has been proposed that local density enhancements (waklets) arising in the
 galactic disk connect with each other and make global spiral arms.
However, the understanding of this mechanism is not yet sufficient.
We analyze the interaction of wakelets by using $N$-body simulations including perturbing point
 masses, which are heavier than individual $N$-body particles and act as the seeds for wakelets.
Our simulation facilitates more straightforward interpretation of numerical results than previous
 work by putting a certain number of perturbers in a well-motivated configuration.
We detected a clear sign of non-linear interaction between wakelets, which make global spiral arms
 by connecting two adjacent wakelets.
We found that the wave number of the strongest non-linear interaction depends on galactic disk mass
 and shear rate.
This dependence is consistent with the prediction of swing amplification mechanism and other
 previous results.
Our results provide unification of previous results which seemed not consistent with each other.

\end{abstract}




\section{Introduction}

The present work focuses on the formation of spiral arms in disk galaxies.
Many researchers have investigated the formation and evolution of spiral arms by using analytic
 methods and numerical simulations.
Despite these efforts, understanding of the formation mechanism of spiral arms is not yet complete.

The first major approach for this topic is the density waves theory advocated by
 \cite{lindblad_1960,lin_shu_1964}.
This theory suggests that spiral structure is not material arm but a quasi-stationary density
 {\it wave} which propagates through the galactic disk with a constant pattern speed not depending
 on galactocentric radius.
It is easy for this theory to explain the long-lived spiral structure, although its origin cannot be
 answered.

Other possibility is the swing amplification advocated by \cite{toomre_1981}.
Swing amplification theory suggests that spiral arms evolve when density enhanced structure is wound
 up by differential rotation of the galactic disk \citep{goldreich_1965, julian_toomre_1966,
 toomre_1981}.
Key of this mechanism is that the rotational direction of epicycle motion coincides with that of
 winding spiral.
By this coincidence, stars in spiral arms remain in high density regions for a long time, and are
 influenced strongly by gravitational force from spiral arms.
This effect causes the rapid growth of spiral arms.

Maximum amplification of density enhancement by this mechanism is examined by previous work.
\cite{toomre_1981} analyzed the maximum amplification factor as a function of $X$ and $Q$ for three
 disk models.
$X$ is the ratio of azimuthal wavelength of spiral pattern and critical wavelength for local
 instability,
\begin{equation}
  X = \frac{\lambda_\phi}{\lambda_c},
\end{equation}
\begin{equation}
  \lambda_c = \frac{4\pi^2G\Sigma}{\kappa^2},
\end{equation}
where $\Sigma$ and $\kappa$ denote surface density and epicycle frequency.
$Q$ is the local disk instability criterion \citep{toomre_1964}.
\cite{toomre_1981} showed that spiral arms are most developed when $X \sim 2$.

\cite{dobbs_baba_2014} performed a similar calculation which builds on \cite{toomre_1981,
 athanassoula_1984}.
They added shear rate of the galactic disk as a new parameter, and showed that spiral arms develop
 at smaller $X$ when shear rate is smaller (see Figure 5 of \citet{dobbs_baba_2014}).

Both simulated and observed galaxies show tendency that the pattern of spiral arms depends on galaxy
 disk property, disk mass and shear rate.
\cite{carlberg_freedman_1985} found that the wave number of the most developed spiral arms is
 inversely proportional to the disk to total mass ratio.
They discussed that this relation results from the existence of characteristic wavelength with
 $X \sim 2$.
\cite{donghia_2015} showed that there is an agreement between the analytic prediction and the
 simulations in term of the number of spiral arms according to swing amplification.
\cite{seigar_2005, seigar_2006} derived the relation of pitch angle and shear rate from observed
 galaxies, while \cite{grand_2013, michikoshi_2014} found a similar relation from simulated
 galaxies, with stronger shear corresponding to more tightly wound arms.

Density wave theory tries to explain spiral arms as long-standing quasi-static structures, with
 their amplitude and pattern speed nearly constant with time.
However, recently performed $N$-body simulations suggest a fundamentally different view.
These work showed that spiral arms are transient structure and alternate between formation and decay
 \citep{carlberg_freedman_1985,bottema_2003,sellwood_2011,fujii_2011,grand_2012a,grand_2012b,
 baba_2013,donghia_2013,roca-fabrega_2013}.
This picture is partly similar to swing amplification theory, but some features of the recently
 simulated arms cannot be explained in terms of swing amplification.
For example, some work proposed that spiral arms co-rotate with stars at each radius
 \citep{wada_2011,grand_2012a,baba_2013,donghia_2013,roca-fabrega_2013}.
So swing amplification cannot grasp all the aspects of spiral arm formation in simulated galaxies.

Another approach was taken by \cite{donghia_2013}, which adds further complexity to the formation
 mechanism of spiral arms.
Their simulation contains many perturbers in low mass $N$-body stellar disks.
Perturbers are modelled as point masses, each of which is heavier than individual $N$-body
 particles.
Their gravity induces local density enhancement around each perturber which they call `wekelet'.
Their proposition is that global spiral arms are formed by connections of wakelets.
Their simulation, however, hampers deeper understanding of the spiral arm formation.
Each simulation contains 1000 perturbers distributed
 with the same profile as the disk and assumed to be corotating on circular orbits.
It is difficult to isolate interaction of a certain pair of perturbers and analyze connection
 process of wakelets because of this complexity.

The purpose of our study is to overcome this difficulty and clarify the fundamental mechanism of
 connection of wakelets.
We performed $N$-body simulation with perturbers used in a similar way to \cite{donghia_2013}.
However our simulation is better controlled. We introduce a smaller number of perturbers, and
 arrange them regularly in a pair of concentric rings in the disk plane.
This setup enables isolating each connection process of wakelets, leading to a more straightforward
 interpretation of numerical results.

We found that two wakelets, which are orbiting at different galactocentric radii, interact
 non-linearly with each other when the inner wakelet overtakes the outer one rotating more slowly
 around the galactic center.
Density enhancement caused by this non-linear interaction connects the two wakelets, thereby form a
 longer density enhancement.
Successive operation of such interaction is considered to create a global spiral arm extending over
 the entire disk.
The wave number of spiral arms developed by this mechanism is found to be consistent with the
 prediction of the swing amplification.
We also investigated how the strength of wakelet connection depends on disk mass and shear rate.
The results can be naturally understood by invoking swing amplification.

The rest of this paper is organized as follows.
In section 2, we describe simulation models and the numerical methods.
We show our results and analyze effect of non-linear interaction in section 3.
In section 4, we show the dependence on disk mass and shear rate.
Finally, we discuss the role of non-linear interaction in the formation and evolution of spiral arms
 in section 5.


\section{Simulations}

\subsection{galaxy models}
In our simulation, each model galaxy consists of a static dark matter halo and a three-dimensional
 $N$-body exponential disk.
Several work shows that the characteristic properties of spiral arms depend on disk mass and shear
 rate as mentioned in introduction.
Therefore, we introduce these two parameters in our disk models.

Shear rate is defined as
\begin{equation}
  \Gamma = \frac{1}{2} \left( 1-\frac{R}{V_c}\frac{dV_c}{dR} \right),
  \label{eq:Gamma}
\end{equation}
 where $V_c$ is rotation velocity at radius $R$.
In our model, rotation curve is prescribed so that shear rate is constant for all radii.
In this case, rotation velocity becomes $V_c \propto R^{-C_{shear}}$, where
\begin{equation}
  C_{shear} \equiv 2\Gamma-1.
  \label{eq:Cshear}
\end{equation}
According to \cite{seigar_2005}, shear rate of real galaxies is $0.2<\Gamma<0.8$,
 so $-0.6<C_{shear}<0.6$.
For fiducial case that $V_c=200{\rm km s^{-1}}$ at $R=8{\rm kpc}$,
\begin{equation}
  V_c = 200 \left[ \frac{R}{8{\rm kpc}} \right] ^{-C_{shear}}{\rm km s^{-1}}.
\end{equation}
We modified this rotation curve to,
\begin{equation}
  V_c = \frac{200}{1+ \left( R_d/R \right)^2 } \left[ \frac{R}{8{\rm kpc}} \right] ^{-C_{shear}}{\rm km s^{-1}},
  \label{eq:V_c}
\end{equation}
 where $R_d$ is the factor which prevents rotation velocity from diverging at center when
 $C_{shear}>0$.

We take an exponential profile as density profile of the stellar disk component.
Rotation velocity arising from disk gravity is then
\begin{equation}
  V^2_{c,disk}=\frac{2GM_{disk}y^2}{R_s} \left[ I_0(y)K_0(y)-I_1(y)K_1(y) \right],
\end{equation}
\begin{equation}
  y \equiv \frac{R}{2R_s},
\end{equation}
 where
 $R_s$ is scale radius and
 I and K are modified Bessel functions \citep{gala_dy_2}.
Here, we assumed that $R_s=3{\rm kpc}$.
Strictly speaking, this velocity profile is for an infinitesimally thin 2D exponential disk.
We use this profile for our 3D simulation, because it is not necessary to set up a exact equilibrium
 initially.
Actually we evolve an isolated model for some time before we introduce perturbers, and make the disk
 relax into equilibrium in a practical sense.

We use five models that have different disk masses or shear rates.
Parameters of each model are listed on Table \ref{tab:model}.
\begin{table}[tbp]
  \begin{center}
    \caption{galaxy models}
    \begin{tabular}{|c|c|c|} \hline
      model & $C_{shear}$ & $M_{disk}$ \\ \hline \hline
      standard model & 0.0 & $1.5\times10^{10}M_{\odot}$ \\
      low mass model & 0.0 & $0.8\times10^{10}M_{\odot}$ \\
      high mass model & 0.0 & $3.0\times10^{10}M_{\odot}$ \\
      weak shear model & -0.3 & $1.5\times10^{10}M_{\odot}$ \\
      strong shear model & 0.3 & $1.5\times10^{10}M_{\odot}$ \\ \hline
    \end{tabular}
    \label{tab:model}
  \end{center}
\end{table}
First, we show results for the standard model in section 3.
In section 4, we describe other models and discuss parameter dependence.

\begin{figure}[tbp]
  \begin{center}
    \includegraphics[width=7cm]{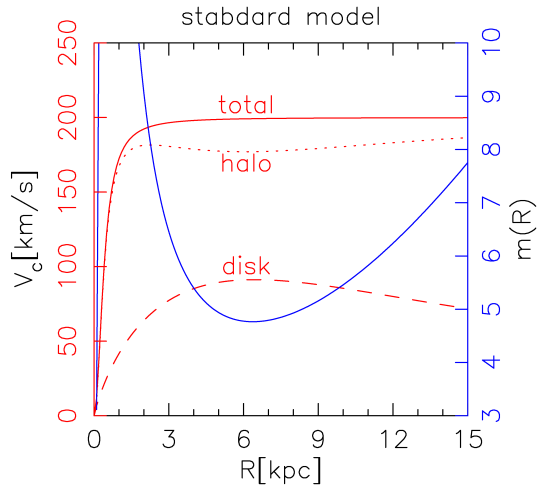}
  \end{center}
  \caption{{\footnotesize
Circular velocity for the standard model. Red solid line shows circular velocity $V_c$.
Red dotted and dashed lines show the circular velocity contributed by the halo and disk,
 respectively.
Blue solid line is the number of spiral arms as a function of radius expected by equation
 (\ref{eq:mr}).}
}
  \label{fig:rot_st}
\end{figure}

$V_c$ and $V_{c,disk}$ are calculated once we specify the $C_{shear}$ and $M_{disk}$ .
Then the rotation velocity caused by the dark matter halo is given by
\begin{equation}
  V^2_{c,halo}=V^2_c-V^2_{c,disk}.
\end{equation}
These velocities for the standard model are shown in Figure \ref{fig:rot_st}.

The disk to total mass ratio $f_d$ is the important parameter because this parameter is strongly
 correlated with the number of spiral arms \citep{carlberg_freedman_1985}.
$m(R)$, which is the number of spiral arms as a function of radius, is expected to be given by
\begin{equation}
  m(R) \sim \frac{1}{f_d(R)} \sim \left( \frac{V_c(R)}{V_{c,disk}(R)} \right)^2,
  \label{eq:mr}
\end{equation}
and shown in Figure \ref{fig:rot_st}.

We model only the disk component as an $N$-body system ($N = 3 \times 10^5$), and treat the halo as
 a static gravitational field.
Initial velocities of disk stellar particles are determined by solving Jeans' equation following
 \cite{hernquist_1993}.

\subsection{numerical method}

We use the GRAPE system of National Astronomical Observatory of Japan for numerical computation.

\begin{figure}[tbp]
  \begin{center}
    \includegraphics[width=7cm]{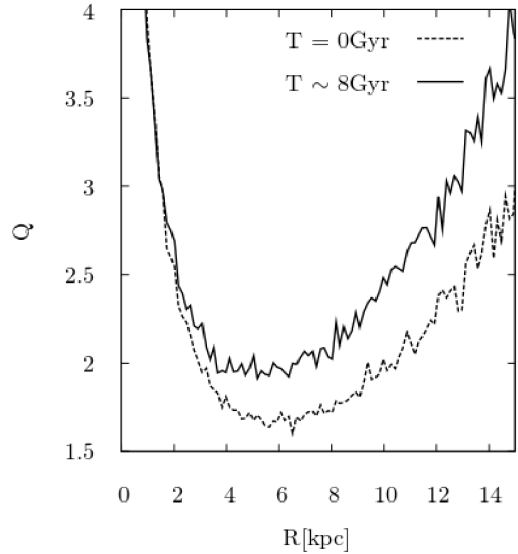}
  \end{center}
  \caption{{\footnotesize
Development of $Q$ value during evolution without perturbers.
Dotted line show initial $Q$ value. Solid line show $Q$ value at $T \sim 8{\rm Gyr}$.
The minimum value of $Q$ increased to about 2.}}
  \label{fig:Q}
\end{figure}

First, we make up a stable disk by using a high $Q$ value, $Q_{min}=1.7$.
We evolved the system for about 8Gyr from this state to remove initial fluctuations which arise
 because the disk is not in a rigorous dynamical equilibrium initially.
Figure \ref{fig:Q} show the development of $Q$ value for this phase.
Eventually, the minimum value of $Q$ increases to about 2.
The final disk does not have any significant spiral structures, and we take it as a stable disk.

We added perturbers to this stable disk in the next step.
Perturbers are expected to play a role as the seed to form the wakelets in our numerical
 experiments.
The mass of each perturber is $5.0\times10^{7}M_{\odot}$ (about 0.3\% of disk mass), and this choice
does NOT have any astrophysical ground (e.g. giant molecular clouds).
We discuss about  possible astronomical origins of wakelets in section 5.2.
In order to isolate dynamical behavior of perturbers, arrangement of perturbers is especially
 important.

Perturber are placed equally spaced in a circle or two circles around the disk center.
We calculated four models as follows.
Models A and B are single ring models.
The former has perturbers located at 6kpc while the latter at 9kpc.
Model C is a double-ring model in which perturber are located at 6kpc and 9kpc (see Figure
 \ref{fig:ss}).
The number of perturbers at each circle was varied from 3 to 10.
In these models, each perturber is made to move on a circular orbit with the same velocity as the
 initial stellar rotational velocity at the same radius.
Namely, pertubers co-rotate with the equilibrium stellar disk.
Additionally, we calculated Model D which has no perturber.
First, we show the case in which 6 perturbers are placed in each ring.


\section{Results}

\begin{figure*}[tbp]
  \begin{center}
    \includegraphics[width=0.96\textwidth]{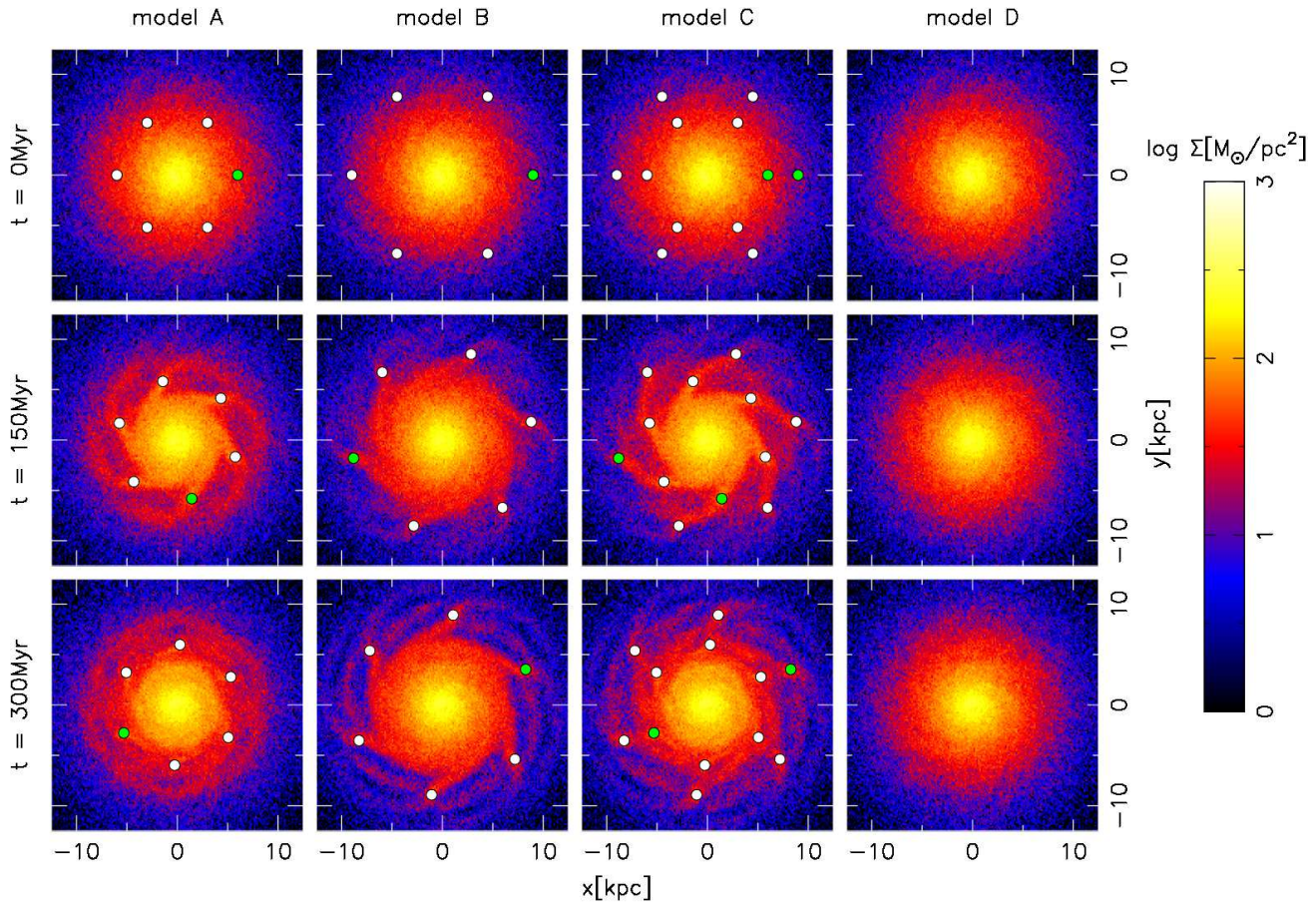}
  \end{center}
  \caption{{\footnotesize
Surface density evolution in four simulations for the standard model.
These results are the case in which the number of perturbers at each circle is 6.
Each column shows Models A, B, C, and D from left to right.
Each row shows three epochs t= 0Myr, 150Myr and 300Myr from top to bottom.
Time is reckoned from the moment when the pertubers were introduced in the simulation.
Large dots indicate the place of perturbers.
Green dots show the perturbers which were located on the x-axis($x>0$ and $y=0$) at $T=0{\rm Myr}$,
 indicating disk rotation at each radius.}}
  \label{fig:ss}
\end{figure*}

Figure \ref{fig:ss} shows snapshots of the surface density distribution for the standard model.
Models A, B, C, and D are displayed from left to right.
In models with perurbers, 6 perturbers are placed in each ring.
For each model, density distribution is shown at three different times from top to bottom.
We added perurbers at $t=0$.

Large dots indicate the location of perturbers at each time, with green points showing the
 perturbers which were located at $x>0$ and $y=0$ at $T=0{\rm Myr}$.

It is noticed that a density enhancement appears around each perturber.
These structures can be considered as wakelets.
As the disk evolves, further density enhancements are created by interaction of two wakelets.
For example, we can see that global spiral arms are formed by conection of wakelets in model C.

\begin{figure}[tbp]
  \begin{center}
    \includegraphics[width=0.42\textwidth]{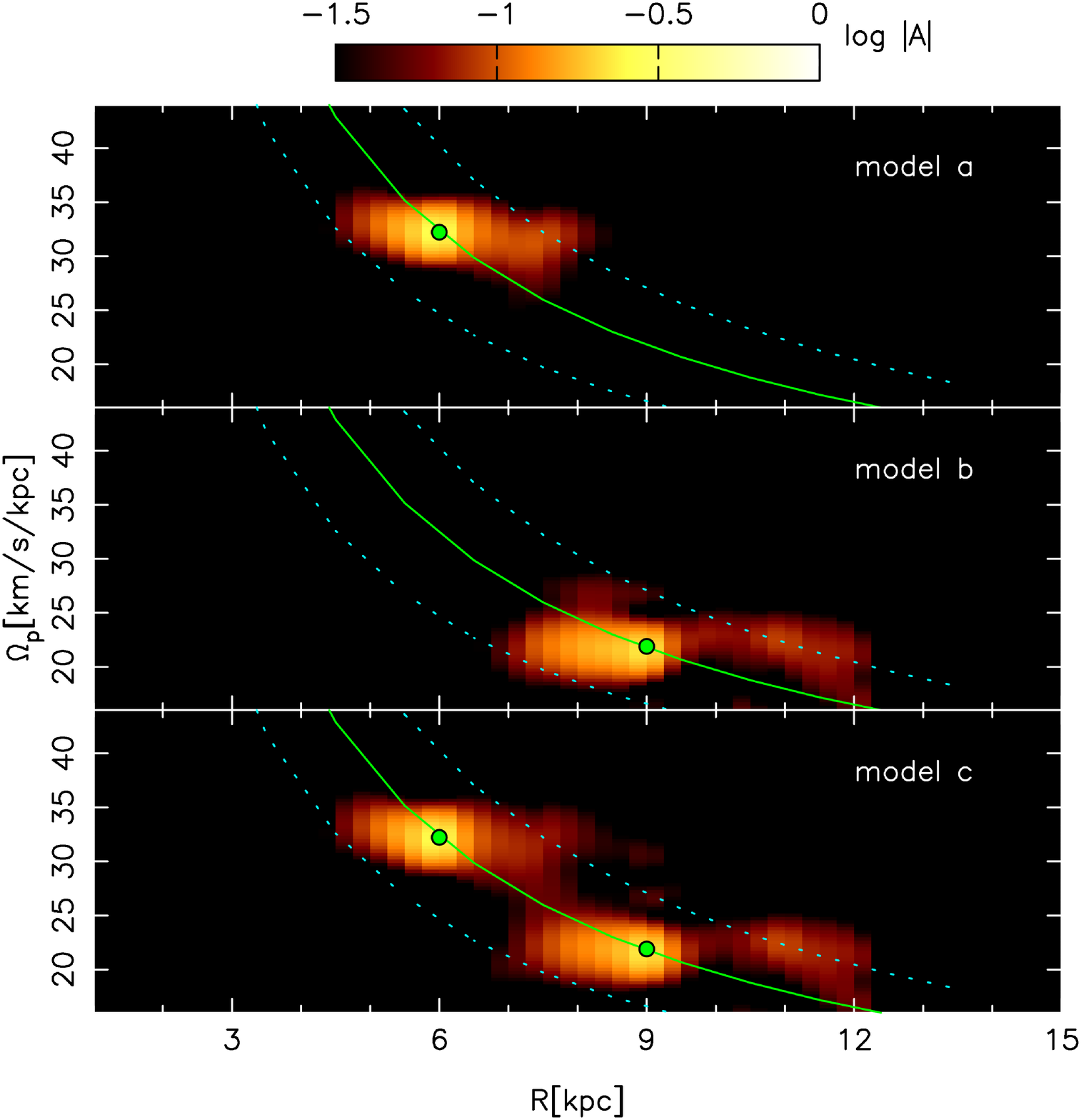}
  \end{center}
  \caption{{\footnotesize
Results of Fourier transformation for models A, B, and C.
Solid green and dotted cyan lines show ($\Omega$, $\Omega \pm \kappa/m$), thus indicating
 co-rotation and Lindblad resonances for a given pattern speed.
Green dots show the galactocentric radius and angular velocity of perturbers.}}
  \label{fig:Fourier}
\end{figure}

We used Fourier transform,
\begin{equation}
  A=\frac{1}{N_f}\sum_k \sum_j exp \left[ im( \phi _{j,t_k}- \Omega_p t_k) \right]
  \label{eq:Fourier}
\end{equation}
 to analyze the pattern speed of spiral arms.
$A$, $N_f$, $\phi_{j,t_k}$ and $\Omega_p$ are Fourier amplitude, the particle number used for
 Fourier transform, azimuth angle of j-th particle at $t=t_k$ and the pattern speed, respectively.
We divided the disk into a series of concentric annuli, and the particles located in each annulus
 were used for Fourier transformation.
We thus get the Fourier amplitude and the pattern speed as functions of the radius.
The width of each annulus is 0.25kpc, and time span of window function is
 $0{\rm Myr}<t<300{\rm Myr}$.

Figure \ref{fig:Fourier} shows the results of this Fourier transform for Models A, B and C.
It is shown that density enhancement is made around each perturber indicated by green dots.
Namely, pertubers form wakelets around themselves also in Fourier space.
It is also clear that each wakeket has a radial extension and has roughly constant pattern speed.
Figure \ref{fig:Fourier} shows that the radial extent of each wakelet is limited by inner and outer
 Lindbrad resonances.
This feature is similar to that of local modes described by \cite{sellwood_carlberg_2014}.
The relation of our wakelets and those local modes is discussed in Section 5.

We are interested in a possible ``non-linear interaction of wakelets".
Here, what ``non-linear" means is the effect other than linear superposition of effects of inner and
 outer perturbers.
We use the phrase ``non-linear interaction of wakelets" in order to emphasize that interaction of
 wakelets causes the density development more than linear superposition of wakelets.
We devised the following method for picking up the effects of non-linear interaction.

We combine the distribution of particles for the four models (Models A, B, C or D) as follows.
Superimposition of Models A and B would have simply effects of inner and outer perturbers linearly
 combined.
On the other hand, superimposition of Models C and D would have additional effect of the non-linear
 interaction between the inner and outer wakelets.
Therefore, the ``enhanced surface density" profile,
\begin{equation}
  \delta \Sigma = ( \Sigma_C + \Sigma_D )-( \Sigma_A + \Sigma_B)
  \label{eq:dSigma}
\end{equation}
 gives the enhancement by non-linear interaction.
In other words, if wakelets do not interact each other, equation (\ref{eq:dSigma}) shows that
 $\delta \Sigma \sim 0$ because Model C is linear superimposition of effects of inner and outer
 perturbers in that case.

\begin{figure}[tbp]
  \begin{center}
    \includegraphics[width=0.48\textwidth]{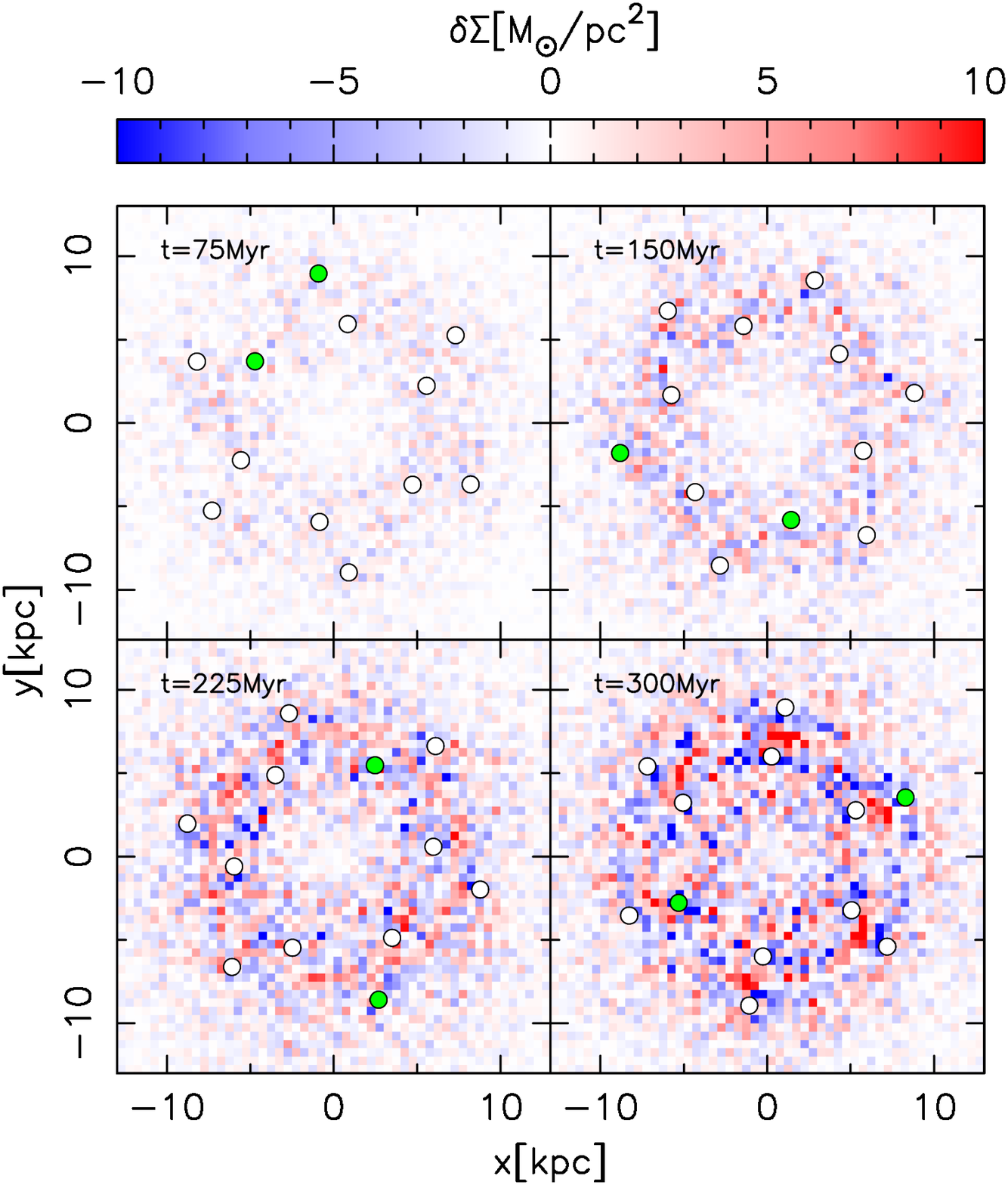}
  \end{center}
  \caption{{\footnotesize
Enhanced surface density profile, equation (\ref{eq:dSigma}), for case of $m=6$.
Color coding shows the $\delta\Sigma$.
Redden regions express the density enhancement by non-linear interaction.
Large dots indicate the place of perturbers.
Green dots show the perturbers which were located at $x>0$ and $y=0$ at $T=0{\rm Myr}$, indicating
 disk rotation at each radius.}}
  \label{fig:dSigma}
\end{figure}

We calculate the enhanced surface density at each time step.
Figure \ref{fig:dSigma} show the results for the case when the number of perturbers at each circle
 is 6.
It is noticed that non-linear interaction makes a density enhancement between the locations of inner
 and outer perturbers in particular as shown at $t=300{\rm Myr}$.
This enhancement connects the inner and outer wakelets temporarily.

\begin{figure}[tbp]
  \begin{center}
    \includegraphics[width=0.48\textwidth]{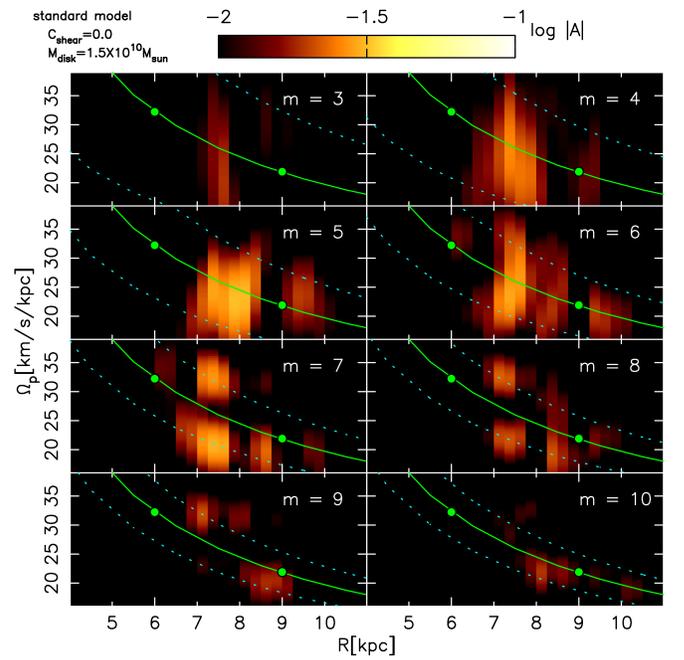}
  \end{center}
  \caption{{\footnotesize
Results of Fourier transformation for non-linear density enhancement; equation (\ref{eq:dSigma}).
Each panel shows results for the case when the number of perturbers is 3 to 10.
The wave number $m$ for Fourier transformation is equal to the number of perturbers.
Solid green line , dotted cyan line and green dots have same meaning as in figure
 \ref{fig:Fourier}.}}
  \label{fig:ft_st}
\end{figure}

Fourier transform of this ``enhanced surface density" gives the pattern speed of non-linear enhanced
 structure.
In order to see the effect of varying the number of perturbers, we carried out the simulations
 placing 3 - 10 perturbers at each ring.
The aim of this numerical experiment is investigating the most developed wave number by non-linear
 interaction.

The number of spiral arms formed by non-linear interaction is equal to the number of perturbers at
 each circle.
Figure \ref{fig:ft_st} shows Fourier amplitude for non-linear enhancement as seen in Figure
 \ref{fig:dSigma} when the number of perturbers at each circle is varied from 3 to 10.
The width of time window function for Fourier transform is 100Myr centered around $t=300{\rm Myr}$.
These results are obtained by calculating equation (\ref{eq:Fourier}) for $\delta \Sigma$.
We use the number of perturbers at each circle as the azimuthal wave number $m$ for Fourier
 transform; equation (\ref{eq:Fourier}).

When the number of perturbers at each circle is 6 ($m=6$), the figure \ref{fig:ft_st} shows a strong
 amplification between 6kpc and 9kpc.
This amplification suggests the presence of a non-linearly developed structure.
More specifically, global spiral arms of Model C as shown in figure \ref{fig:ss} include not
 only simple linear superposition of inner and outer wakelets but also non-linear interaction of
 those wakelets.
It is also noted that this structure has a pattern speed which is between those of the two
 perturbers.
Comparing the results for different $m$ suggests that spiral arms are most developed by non-linear
 interaction when there are 5 or 6 perturbers at each radius for our standard model.
It is noted that these wave numbers ($m=5,6$) are consistent with $m(R)$ between $6{\rm kpc}$ and
 $9{\rm kpc}$ in Figure \ref{fig:rot_st}.
It is very interesting that non-linear interaction does not always develop when two wakelets
 encounter.
But this result does NOT mean that the number of spiral arms depends on the number of perturbers.
Figure \ref{fig:ft_st} only shows that global spiral arms whose wave number is 5 or 6 are selectively
 developed by non-linear interaction, while spiral arms with other wave numbers grow inefficiently.
Thus the number of global spiral arms developed by non-linear interaction is 5 or 6
 independently of the number of perturbers or wakelets.
This prediction is consistent with the results of \cite{donghia_2013} that the number of spiral arms
 does not depends on the number of perturbers.

These non-linear enhancements by interaction of wakelets are suggestive of the non-linear mode
 coupling theory suggested by \cite{tagger_1987} and
the presence of mode coupling between bar and spiral arms demonstrated by some simulations
 (e.g. \cite{masset_1997, quillen_2011}).
Spiral arms may be the product of mode coupling of wakelets.


\section{Other Galaxy Models}

We also simulated other disk models, having different mass or shear rate.
We investigated the relationship of the wave number and amplitude of non-linear enhancements
 for these disks in the same way as for the standard model, and found out the wave number of the
 most developed structure for each disk.
Hereafter we refer to this wave number as the characteristic wave number.

\begin{figure}[tbp]
  \begin{center}
    \includegraphics[width=0.48\textwidth]{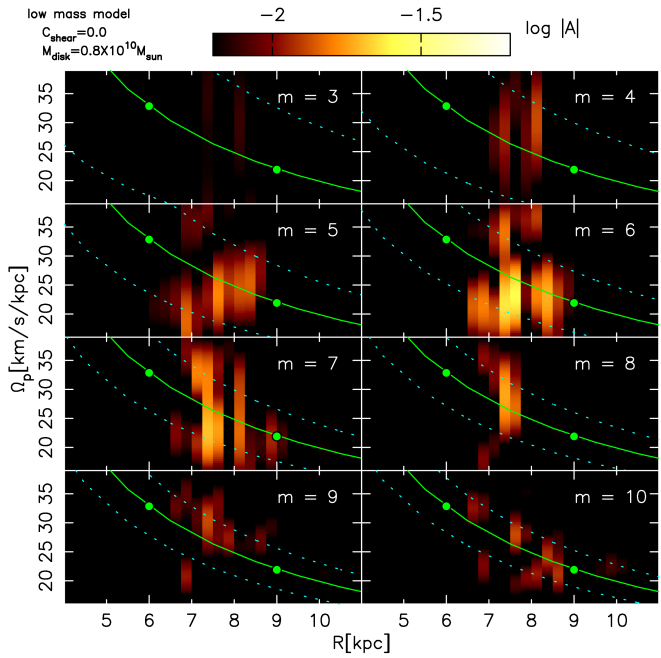}
  \end{center}
  \caption{{\footnotesize
Same as Figure \ref{fig:ft_st}, but for low mass models.}}
  \label{fig:ft_lm}
\end{figure}

\begin{figure}[tbp]
  \begin{center}
    \includegraphics[width=0.48\textwidth]{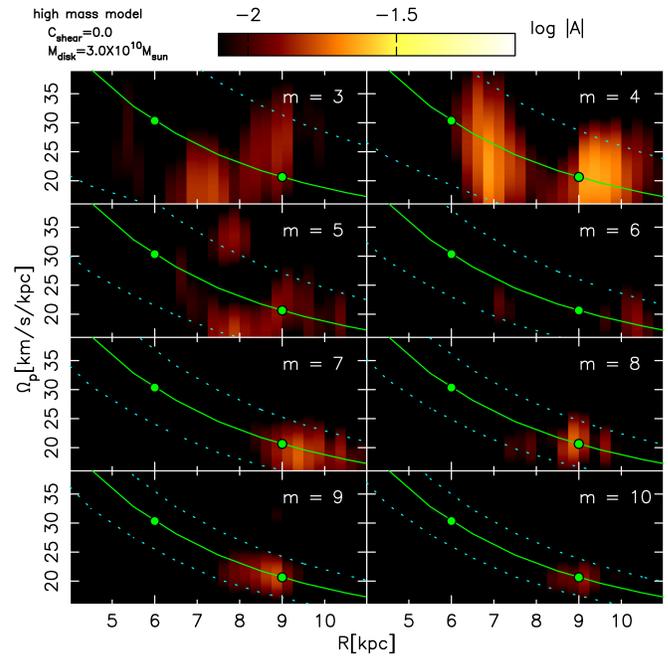}
  \end{center}
  \caption{{\footnotesize
Same as Figure \ref{fig:ft_st}, but for high mass models.}}
  \label{fig:ft_hm}
\end{figure}

First, we simulated higher and lower mass disk models than the standard model.
Figure \ref{fig:ft_lm} and \ref{fig:ft_hm} is the results of Fourier transform for ``enhanced
 surface density" of low mass and high mass models.
Comparing three models, the standard, the high-mass and the low-mass models, shows that
 the characteristic wave number depends on the disk mass.
It is clear that the lower mass disk develops spiral arms with larger wave number as a result of
 non-linear interaction of wakelets.
For the low-mass model, non-linear enhancement is the strongest at $m=6-8$, whereas for the
 high-mass model non-linear enhancement is the strongest at $m=4$.
This dependence is consistent with the tendency observed in $N$-body simulations of disk galaxies
 \citep{carlberg_freedman_1985,bottema_2003}.


\begin{figure}[tbp]
  \begin{center}
    \includegraphics[width=0.48\textwidth]{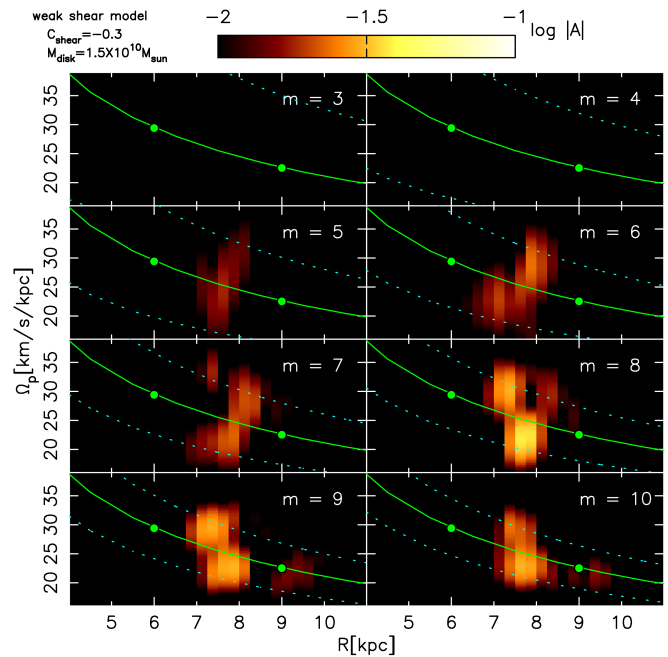}
  \end{center}
  \caption{{\footnotesize
Same as Figure \ref{fig:ft_st}, but for weak shear models.}}
  \label{fig:ft_ws}
\end{figure}
\begin{figure}[tbp]
  \begin{center}
    \includegraphics[width=0.48\textwidth]{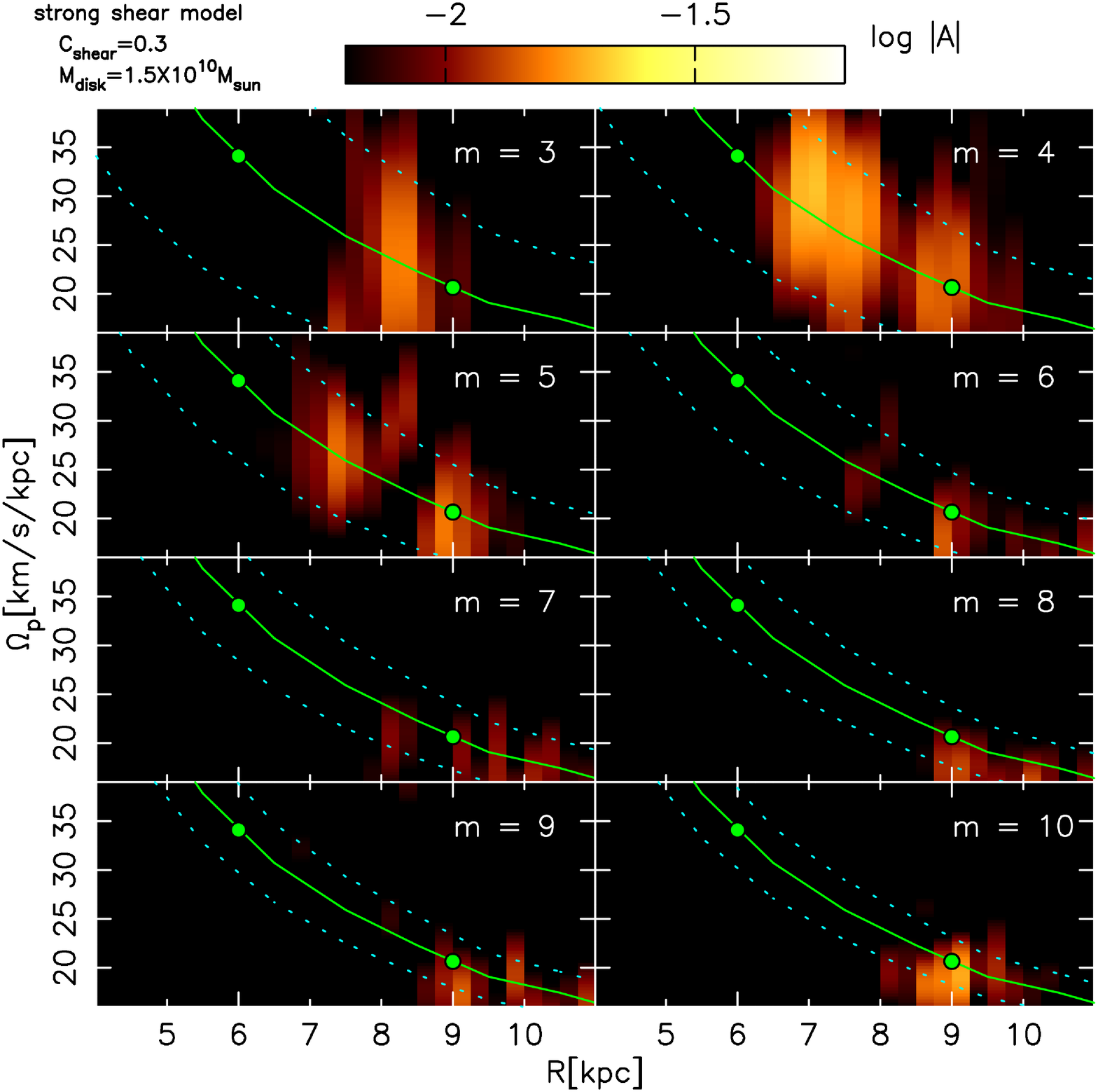}
  \end{center}
  \caption{{\footnotesize
Same as Figure \ref{fig:ft_st}, but for strong shear models.}}
  \label{fig:ft_ss}
\end{figure}

Next, we carried out simulations introducing stronger and weaker shear in the disk.
Figure \ref{fig:ft_ws} and \ref{fig:ft_ss} give the results of Fourier transform for ``enhanced
 surface density" in the weak-shear and strong-shear models.
These results show that weaker shear leads to a smaller number of spiral arms.
We discuss this dependence on shear rate in section 5.

To summarize, we found that the characteristic wave number of spiral arms depends on the disk
 properties, disk mass or shear rate.
A disk with lower mass or weaker shear develops smaller number of spiral arms by non-linear
 interaction of wakelets.
%
%
\section{Discussion}
Our simulation clearly detected non-linear interaction of wakelets originally caused by perturbers
 rotating in the disk.
It was also demonstrated that this non-linear interaction plays a fundamental role in the formation
 of global spiral arms.
We here discuss relevance of these findings to the previous results and try to interpret numerical
 results reported by other authors which seemingly contradict each other.
\subsection{the dependence on mass and shear}
We found that the strength of non-linear interaction depends on disk mass and shear rate.
This dependence may be explained by the swing amplification theory.
Swing amplification takes place when the density enhancement winds up due to disk differential
 rotation.
In our simulation, wakelets formed around perturbers connect with other and a region of excess
 density is formed which extends spatially from the inner to outer perturbers.
This region winds up since the two perturbers, which move at different radius from each other, have
 different rotational speeds.
Therefore, the structure formed by connection of wakelets will be subject to swing amplification.
It is known that the efficiency of swing amplification depends on disk properties.
\cite{carlberg_freedman_1985} showed that the wave number of spiral arms developed most by swing
 amplification is inversely proportional to the disk mass fraction, namely, the disk mass divided by
 the total mass of the galaxy.
On the other hand, \cite{dobbs_baba_2014} showed that the $X$ parameter, which measures the
 efficiency of swing amplification, varies with the shear rate.
When the shear rate is strong (weak), swing amplification effectively operates for a large (small)
 $X$.
These trends are consistent with our results.
This correspondence gives more support to our hypothesis that the structure formed by connection of
 wakelets experience swing amplification.
\subsection{Ubiquity of non-linear interaction}
Our simulations have several artificial settings.
For example high $Q$ value is adopted for which spontaneous spiral formation is largely prohibited.
In order to create density enhancements in these stable disks, point mass perturbers are employed.
Then there arises a natural questions, ``Does non-linear interaction occur in unstable $N$-body disk
 without perturbers?"
Our answer is ``Yes." because of following reasons.

Our simulations show that the lighter the disk is, the larger the characteristic wave number of
 spiral arms developed by non-linear interaction of wakelets becomes.
Previous work about unstable $N$-body disks without perturbers shows a similar trend that the
 lighter disk is, the larger the wave number of emergent spiral arms becomes.
This correspondence makes us anticipate that unstable $N$-body disks also develop spiral arms by
 non-linear interaction of wakelets.

The non-linear interaction mechanism in our study needs perturbers as the seed of wakelets.
However, many previous work show that spiral arms are formed in pure $N$-body disk without perturbers.
What would play a role of perturbers in pure $N$-body disks?
\cite{sellwood_carlberg_2014} indicate the presence of local wave modes in their
 simulation which deals with pure $N$-body disk.
\cite{donghia_2013} showed that after removing the perturbers from disk the wakelets survive owing to sufficiently
 high density to serve as perturbers themselves and the stellar disk holds up spiral activity.
It is therefore possible that density enhancements like local modes are also able to act as perturbers or
 wakelets and cause non-linear interaction.
That is to say, wakelets may be formed spontaneously by the disk self-gravity without perturbers in the essentially same manner as
 local modes are formed in pure $N$-body disk.
In support of this conjecture, the pattern speeds of wakelets in our models (figure \ref{fig:Fourier}) are similar to those of the
 local modes found by \cite{sellwood_carlberg_2014}.
It is stressed that each local mode extends radially and its pattern speed is constant along its
 extent, essential features of our wakelets.
In summary, our models required perturbers to induce wakelets because they are stable by
 construction and do not create any density enhancements spontaneously.
In unstable disks, local density perturbations created by the self gravity of the disk in early
 evolution phase (such as the local modes of \cite{sellwood_carlberg_2014}) will serve as wakelets,
 and non-linear interactions form global spiral arms by connecting neighboring density enhancements.

\subsection{co-rotation wave and local mode}

\cite{sellwood_carlberg_2014,mata-chavez_2014} showed existence of local wave mode having a certain
 pattern speed.
On the other hand, some recent work indicates that spiral arms co-rotate with stellar particles at
 each radius \citep{wada_2011,grand_2012a,baba_2013,donghia_2013,roca-fabrega_2013}.
How can these two results be compatible with each other?

Non-linear interaction mechanism can link these two views.
In our simulation, non-linearly enhanced structures have the pattern speed between those of the
 inner and outer perturbers.
It is also located spatially between the two perturbers which co-rotate with disk stars.
Therefore the structure developed by non-linear interaction inevitably co-rotate with disk stars.
Global spiral arms are made by successive operation of non-linear interactions of wakelets (or local
 modes of \cite{sellwood_carlberg_2014}.
This gives a natural explanation why global spiral arms manifest as co-rotation waves.

Why did \cite{sellwood_carlberg_2014} see their spiral arms as distinct Fourier components (local
 `modes')?
We can also answer to this question.
The effect of non-linear interaction alternates between strong and weak phases.
This is an essential feature of swing amplification.
Therefore, when the window function for Fourier transformation has a wide range in time domain,
 non-linear effect is attenuated because it include weak phases as well as strong phases.
If we calculate Fourier amplitudes for only strong phases, the results show the presence of
 structure co-rotating with stellar particles at each radius.
Fourier analysis with a wide time range shows local modes whereas adoption of a narrow time range
 shows co-rotation waves.

\subsection{other spiral structure}

The results of the present work are purely theoretical.
However, they have an interesting observational implication.
It is seen from Figure \ref{fig:ss} that the spiral arms formed by non-linear interaction of
 wakelets exhibit straight-line structures.
Interestingly enough, observations of real disk galaxies often show straight-line structures in the
 spiral arms \citep{velyaminov_1964,chernin_2000}.
The most clear examples include M101 and M51.
Our mechanism may be able to explain these remarkable structures.

Another notable feature, which may be relevant to our finding, is the branching of spiral arms often
 observed in multi-armed grand-design spiral galaxies.
This may be realized by putting {\it different} number of perurbers at inner and outer radii.
In our simulation, inner and outer rings have the same number of perturbers, so that inner wakelets
 connect with outer counterparts one-to-one.
When the outer ring has more perurbers than the inner one, some inner wakelts may connect to two or
 more outer wakelets, thus bringing about bifurcation of spiral arms.
This interesting possibility deserves further numerical investigation.


\section{Summary}

We analyzed the interaction of wakelets by using $N$-body simulations including perturbing point
 masses, which are heavier than individual $N$-body particles and act as the seeds for wakelets.
Consequently, we got the following results:

\begin{enumerate}

\item
Two adjacent wakelets, which are orbiting at different galactocentric radii, interact non-linearly
 with each other when the inner wakelet overtakes the outer one rotating more slowly around the
 galactic center.
This non-linear interaction make density enhancement and connects the two wakelets, thereby create a
 global spiral arm extending over the entire disk.
(See section 3 and 5.2.)

\item
The wave number of spiral arms developed by this mechanism depends on disk mass and shear rate.
This dependence is consistent with the prediction of the swing amplification and suggests that the
 structure formed by connection of wakelets experoence swing amplification and develop into global
 spiral arms.
(See section 4 and 5.1.)

\item
In our simulation, non-linearly enhanced structures have the pattern speed between those of the
 inner and outer perturbers.
It is also located spatially between the two perturbers which co-rotate with disk stars.
This result provides unification of previous results, namely local wave modes
 \citep{sellwood_carlberg_2014,mata-chavez_2014} and spiral arms co-rotating with stellar particles
 at each radius \citep{wada_2011,grand_2012a,baba_2013,donghia_2013,roca-fabrega_2013}.
(See section 5.3.)

\end{enumerate}

\acknowledgments
Numerical computations in this paper were performed on GRAPE system at Center for Computational
 Astrophysics, National Astronomical Observatory of Japan.

%

%
\end{document}